\documentstyle[sprocl]{article}

\input{psfig}

\def\PRL{\em Phys. Rev. Lett.}

\def\eg{{\it e.g.~}}

\def\be{\begin{equation}}
\def\ee{\end{equation}}
\def\bea{\begin{eqnarray}}
\def\eea{\end{eqnarray}}

\def\ket{{\rangle}}
\def\bra{{\langle}}
\def\ie{{\it i.e.}}
\def\be{\begin{equation}}
\def\ee{\end{equation}}
\def\ba{\begin{eqnarray}}
\def\ea{\end{eqnarray}}
\def\mref#1{Eq. (\ref{Eq:#1})}
\def\mreff#1{Fig. \ref{Eq:#1}}

\def\mlab#1{\label{Eq:#1}}
\def\mlabf#1{\label{Eq:#1}}

\def\half{\frac{1}{2}}
\def\to{\rightarrow}
\def\nn{\nonumber\\}
\def\sk{\vskip 1cm}

\def\mat#1#2#3{\langle{#1}\vert{#2}\vert{#3}\rangle}
\def\etal{{\it et al.}}

\def\PR#1#2#3 {{\it Phys. Rev. }{\bf D#1} #2 {(#3)}}
\def\PRL#1#2#3 {{\it Phys. Rev. Lett. }{\bf #1} #2 {(#3)}}
\def\PL#1#2#3 {{\it Phys. Lett. }{\bf #1} #2 {(#3)}}
\def\AP#1#2#3 {{\it Ann, Phys. }{\bf #1} #2 {(#3)}}
\def\ZP#1#2#3 {{\it Z. Phys. }{\bf #1} #2 {(#3)}}
\def\NP#1#2#3 {{\it Nucl. Phys. }{\bf #1} #2 {(#3)}}
\def\MPL#1#2#3 {{\it Mod. Phys. Lett. }{\bf #1} #2 {(#3)}}
\def\NC#1#2#3 {{\it Nuov. Cim. }{\bf #1} #2 {(#3)}}
\def\PREP#1#2#3 {{\it Phys. Report }{\bf #1} #2 {(#3)}}
\def\PROG#1#2#3 {{\it Prog. Theor. Phys. }{\bf #1} #2 {(#3)}}

\def\cp{{\bf CP}}

\def\sq2{{1\over{\sqrt{2}}}}

\def\g5{\gamma_5}

\renewcommand{\Im}{\mbox{Im}\,}
\renewcommand{\Re}{\mbox{Re}\,}

\begin{document}
\rightline{DPNU-99-12}
\sk
\title{SUMMARY\\
OF\\
WEAK DECAYS, CKM AND CP VIOLATION SESSION\\
PENGUINS}

\author{A. I. SANDA}

\address{Department of Physics, Nagoya University\\
Nagoya 464-01, Japan
\\E-mail: sanda@eken.phys.nagoya-u.ac.jp} 


\maketitle
\abstracts{
In our subgroup, we focused on penguin physics from various different angles. 
Our discussions included: (1) A method to extract one of the angles of the unitarity triangle  $\phi_2$; (2) Methods to extract $\phi_3$ from $B\to K\pi$ decays; (3) Effects  of non-minimal SUSY on B and K decays; (4) Understanding large
branching ratios for $B\to \eta'+K(K^*)$ decays; (5) New calculation for hadronic matrix elements which are needed to compute $\frac{\epsilon'}{\epsilon}$.
}

\section{Introduction}
Our session was given a rather general title: 
"Weak Decays, CKM and CP Violation".
It is a big field and we can not do justice to 
any of these subjects if we try to cover everything. For this reason, 
we decided to concentrate
our discussion on penguin physics.

Year 1998 was a very good year for penguin physics - 
year 1999 promises to be even better for flavor physics in general. 
\begin{enumerate}
\item
We are supplied 
with branching ratios on hadronic two body decays
from CLEO \cite{cleo}:
\ba
{\rm Br}(B^\pm\to K^\pm\pi^0)&=&(1.5\pm 0.4\pm 0.3)\times 10^{-5},\nn
{\rm Br}(B^\pm\to K\pi^\pm)&=&(1.4\pm 0.5\pm 0.2)\times 10^{-5},\nn
{\rm Br}(B\to K^\mp\pi^\pm)&=&(1.4\pm 0.3\pm 0.1)\times 10^{-5},\nn
{\rm Br}(B\to \pi^\mp\pi^\pm)&=&\left(0.37{{+0.20}\atop{-0.17}}\right)\times 10^{-5},\nn
{\rm Br}(B\to \pi^\mp\pi^0)&=&\left(0.59{{+0.32}\atop{-0.27}}\right)\times 10^{-5}.
\mlab{CLEO}
\ea

\item
New results on 
$\frac{\epsilon'}{\epsilon}$ was promised, and in fact, just after the meeting
the result from E832 \cite{E832} was announced. The result is considerably larger than that of previous Fermilab result\cite{E731}:
$${\rm Re} \frac{\epsilon'}{\epsilon} =  \left\{
\begin{array}{ll}
(23\pm 6.5)\times 10^{-4}&{\rm NA31}
\\
(28.0 \pm 0.30_{stat.}\pm 0.26_{syst.}\pm 0.10_{{\rm MC} stat.} )
\times 10^{-4}&{\rm E832}
\end{array}
\right.$$
This result is now in agreement with that from CERN \cite{NA31}
 and establishes non-vanishing direct \cp~violation.
\item Belle and Babar collaborations should start taking data on much 
anticipated large CP violation in B decays. Along the way, there will get lots of data on B decays.
\item 
A new K meson factory at Da$\Phi$ne should start taking data this year.
\end{enumerate}
Who knows, by year 2001, we may have a positive signal for New Physics.
Much of above experimental development demands better theoretical understanding of penguins. 

\section{How big are penguins?}
Let us illustrate the importance of penguins by giving a hand-waving argument
based on the experimental result \mref{CLEO}. Less hand-waving argument is 
presented in Gronau's contribution \cite{gro2}. 
\begin{figure}[t]
\psfig{figure=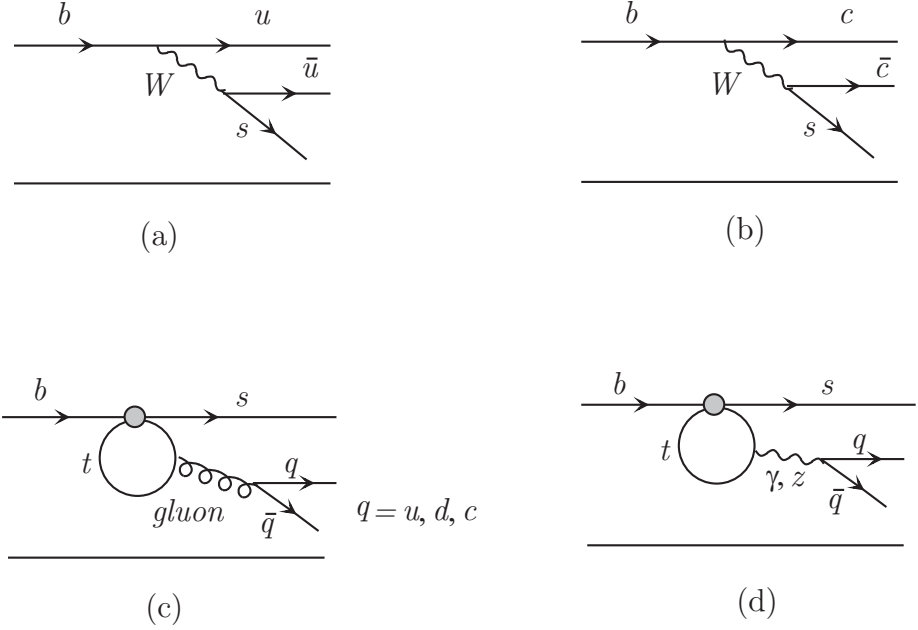,height=3.3in}
\caption{Quark diagrams for 
$B^+\to K^+\pi ^0$ and
$\pi^+\pi ^0$ decays.
(a) The tree graph contribution. 
(b) Penguin contribution.\mlabf{fig1}}
\end{figure}

The amplitudes for tree and penguin contributions for
$K\pi$ decay mode are:
\ba 
T(K\pi)&=& \frac{G_F}{\sqrt{2}}{\bf V}_{ub}^*{\bf V}_{us}
[C_1(\mu)Q_{s1}^u(K\pi)+C_2(\mu)Q_{s2}^u(K\pi)]\nn
P(K\pi)_c&=&\frac{G_F}{\sqrt{2}}{\bf V}_{cb}^*{\bf V}_{cs}
[C_1(\mu)Q_{s1}^c(K\pi)+C_2(\mu)Q_{s2}^c(K\pi)]\nn
P(K\pi)_t&=&\frac{G_F}{\sqrt{2}}(-{\bf V}_{tb}^*{\bf V}_{ts})
\displaystyle\sum_{i=3}^{10}C_i(\mu)Q_{si}(K\pi).
\ea
To simplify our notation, set:
\be
T(K\pi)= {\bf V}_{ub}^*{\bf V}_{us}{\cal T};~
P(K\pi)_c={\bf V}_{cb}^*{\bf V}_{cs}{\cal P}^c;~
P(K\pi)_t=-{\bf V}_{tb}^*{\bf V}_{ts}{\cal P}^t.
\mlab{3}
\ee

If penguin diagram gave negligible contribution, the entire two body decays 
occur through \mreff{fig1}(a). Then $B\to\pi\pi$ decay goes through a diagram in which the $s$ quark in \mreff{fig1}(a) is replaced by a $d$ quark.
For a rough argument, we ignore SU(3) breaking in the hadronic matrix elements.
 Then, $B\to \pi\pi$ is given by
$T(\pi\pi)\sim\lambda^3{\cal T}$. 
Since $T(K\pi)/T(\pi\pi)\sim\lambda$, we would expect:
\be
{\rm Br}(B\to K\pi)/{\rm Br}(B\to \pi\pi)\sim{\cal O}(\lambda^2).
\ee
Experimentally this is not so. This indicates that the $P(K\pi)$ amplitude is at least as large as the $T(\pi\pi)$. 

For $B\to K\pi$,  $P(K\pi)$ is ${\cal O}(\lambda ^2)$  and 
$T(K\pi)$ is ${\cal O}(\lambda ^4)$. So, $P(K\pi)$ must be a major contributor
to the decay amplitude.
If $P(K\pi) \simeq T(\pi\pi)$, then
\be
\frac{{\cal P}^c+{\cal P}^t}{{\cal T}}={\cal O}(\lambda),
\ee
the penguin contribution is considerably larger than 
what a naive estimate of the 
loop graph would suggest: 
\be
\frac{{\cal P}^c+{\cal P}^t}{{\cal T}^u}
\sim
\frac{\alpha _S}{12\pi ^3} 
{\rm log} \frac{m_t}{m_c} \sim {\cal O}(0.01).  
\ee
Gronau went through less hand-waving analysis and obtained
\be
\frac{{\cal P}^c+{\cal P}^t}{{\cal T}^u}
=0.3\pm 0.1.
\mlab{gro}
\ee

\begin{figure}[t]
\psfig{figure=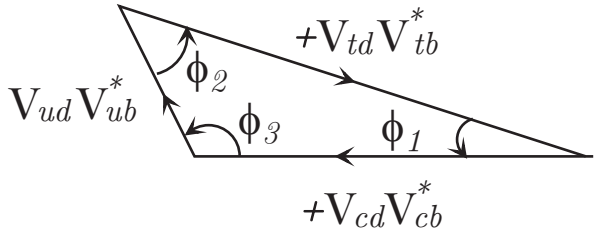}
\caption[]{Angles of the unitarity triangle are related to the phases of the
KM matrix. The right hand rule gives the positive direction of the 
angle between two vectors.}
\mlabf{unitangle}
\end{figure}

Large penguin contribution is not always welcome. For example, they
play a role in so called "penguin pollution" which causes hadronic uncertainty 
in determining $\phi_2$ and $\phi_3$. For notations see \mreff{unitangle} \footnote{ Here we use the notation which was introduced
when the unitarity triangle was first discussed in the context we
use today \cite{srs}.}. Problem penguins cause, however, does not compare
with richness they bring to flavor physics.
Unlike in K decays where effects of tree graphs dominate, in B physics,
 quantum loop effects via penguins is often a leading 
contribution. This gives us a window of opportunity to look for effects beyond the standard 
model - as they are likely to contribute through loop effects.
Anticipating this possibility, we had the following discussions:
\begin{enumerate}
\item Reviews of penguins in B decays by M. Gronau.
\item New remarks on the determination of $\phi_1$ and $\phi_2$ by L. Oliver.
\item A critical look at $\phi_3$ from $B\to K\pi$ by R. Fleischer.
\item Model independent anlysis of $B\to K\pi$ decays and bounds on the weak
phase $\phi_3$ by M. Neubert.
\item Analysis of $B\to\eta K(K^*)$ and $B\to\eta' K(K^*)$ by D-D. Du.
\item Effects of SUSY particles in B and K decays by A. Masiero.
\item Chiral methods and predictions for $K\to\pi\pi$ by E. Paschos.
\end{enumerate}

\section{Taming the penguins}
How to get around the penguin pollution and extract the value of $\phi_2$
has been reviewed by Gronau. Oliver has presented an alternative approach
which may be useful. In his approach, $\phi_2$ is expressed as a function of
experimentally measurable quantities in $B\to\pi\pi$ decay, plus one other 
parameter. It can be, for example $\frac{\cal P}{\cal T}$, obtained from 
$B\to K\pi$, mentioned above. 

The time dependent $B(\overline B)\to\pi^+\pi^-$ decay rates are given by:
\ba 
\Gamma\left({{\overline B}\atop B}\to\pi ^+\pi ^-\right) (t) &\sim& 
\bigg[ |A|^2+ |\overline A| ^2
\mp \left(|A|^2- |\overline A| ^2 \right) \cos (\Delta M_{B_d} t)
\nn
&\pm& 2 {\Im} \left(\frac{q}{p} \overline A A^*\right) 
\sin (\Delta M_{B_d} t) \bigg]. 
\mlab{10.30}
\ea
There are three experimental observables:
$$
|A|,~|\overline A|,~\mbox{and}~{\Im} \left(\frac{q}{p} \overline A A^*\right).
$$
Theoretically, we can write
\ba
A&=&{\bf V}_{ud}{\bf V}_{ub}^*M^u+{\bf V}_{td}{\bf V}^*_{tb}M^t\nn
&=&{\bf V}_{ud}{\bf V}_{ub}^*M^u\left[1-
\left|\frac{{\bf V}_{td}{\bf V}^*_{tb}}{{\bf V}_{ud}{\bf V}_{ub}^*}\right|
e^{i\phi_2}
\frac{M^t}{M^u}\right].
\ea
Here $M^u={\cal T}^u-{\cal P}^c$, and $M^t={\cal P}^t-{\cal P}^c$. These are
related to previously introduced matrix elements except for the SU(3) breaking
corrections.

Theoretical unknowns are 
$$
|M^u|,~\left|\frac{M^t}{M^u}\right|,~\arg\left(\frac{M^t}{M^u}\right),~\phi_2.
$$
Since there are 4 theoretical parameters and only 3 experimental observables,
we can not solve for $\phi_2$. We can, however solve for $\phi_2$ as a function
of, \eg $\left|\frac{M^t}{M^u}\right|$. We can, most likely obtain this 
parameter from \mref{gro} looking at $B\to K\pi$ decays. 
Further study is necessary to
see how the error from SU(3) symmetry breaking will affect the determination of
$\phi_2$. Also, there are some ambiguities coming from the sign of a square root
as well as that coming from $\phi_2\pm\pi$. For details see Oliver's talk
 \cite{charl}.

\section{Getting maximum out of $B\to K\pi,~\pi\pi$ decays}
The CLEO collaboration has recently reported the observation of $B\to K\pi$
decays given in \mref{CLEO}.
It is clearly important to understand what we can learn from these results.
Contributions from Fleischer, Gronau, and Neubert on this subject are rather 
technical. Nevertheless, it is an important technicality, as it must be understood when information is extracted from data. So, rather than summarizing what they have presented, I have presented necessary formalism which I hope is useful
in following their work.

Feynman graphs shown in \mreff{fig1} illustrates the class of operators
which are generated by QCD and electroweak radiative corrections.
The weak Hamiltonian which causes these decays can be written as
\ba
{\cal H}
&=&\frac{G_F}{\sqrt{2}}\biggl\{\xi_u[C_1(\mu)O_1^u+C_2(\mu)O_2^u]
+\xi_c[C_1(\mu)O_1^c+C_2(\mu)O_2^c]\nn
&-&\xi_t\sum_{i=3}^{10}C_i(\mu)O_i\biggr\}
+\hbox{h.c.,}
\mlab{12}
\ea
where $\xi_q={\bf V}^*_{qb}{\bf V}_{qs}$, $O_1^q=(\bar bs)_{V-A}(\bar qq)_{V-A}$ and 
$O_2^q=(\bar bq)_{V-A}(\bar qs)_{V-A}$,
$O_9=\frac{3}{2}(\bar bs)_{V-A}\sum_qe_q(\bar qq)_{V-A}$,
$O_{10}=\frac{3}{2}(\bar bq)_{V-A}\sum_q e_q(\bar qs)_{V-A}$.
Let us try to understand the isospin structure of these operators.
The up tree graph \mreff{fig1}(a) contains  $u$ and $\bar u$ quarks
and they generate both $\Delta I=0$, and $\Delta I=1$ terms in the 
effective Hamiltonian.
\mreff{fig1}(b), the charm tree graph, contains all isosinglet quarks and thus they generate
 $\Delta I=0$ operator. \mreff{fig1}(c), the penguin, gives contribution which is proportional to $\sum_{q=u,d,s,c}\bar q\lambda^aq$ and it gives only $\Delta I=0$ operator.
Finally, \mreff{fig1}(d), the electroweak penguin, gives both  $\Delta I=0$, and $\Delta I=1$ operators.

Now we consider the isospin properties of the up tree, the operator which is generated by \mreff{fig1}(a). Because it contains both
$\Delta I=0$, and $\Delta I=1$ components, we want to separate the operator into two parts:
\ba
2[C_1(\mu)O_1^u+C_2(\mu)O_2^u]&=&C_+(\mu)[O_+^u-O_+^d]+C_-(\mu)[O_-^u-O_-^d]\nn
&+&C_+(\mu)[O_+^u+O_+^d]+C_-(\mu)[O_-^u+O_-^d]
\ea
where $C_\pm=C_2\pm C_1$ and $O_\pm=\half(O_2\pm O_1)$.
Then the first two terms on the right hand side cause $\Delta I=1$ transition and the last two terms cause $\Delta I=0$ transition.

Next we show that the electroweak penguins, can be expressed interms of existing operators \cite{nr,flei}.
Note that the standard model predicts that $C_{7,8}(m_b)$ are very small and 
they are  negligible compared to $C_{9,10}(m_b)$.
The operators with dominant coefficients $O_9$ and $O_{10}$, for $q=u,~d$, 
can be written as linear combinations of
$O_1^{u,d}$ and $O_2^{u,d}$ respectively:
\ba
{\cal H}_0
&=&\frac{G_F}{\sqrt{2}}\biggl\{
\xi_c[C_+(\mu)O_+^c+C_-(\mu)O_-^c]
+
[\xi_uC_+(\mu)-\half\xi_tC_+^{EW}(\mu)]\hat O_+\nn
&+&[\xi_uC_-(\mu)-\frac{1}{2}\xi_tC_-^{EW}(\mu)]\hat O_-
-\xi_t\sum_{i=3}^6C_i(\mu)O_i
\biggr\}\nn
{\cal H}_1
&=&\frac{G_F}{\sqrt{2}}\biggl\{
[\xi_uC_+(\mu)-\frac{3}{2}\xi_tC_+^{EW}(\mu)]\bar O_+
+[\xi_uC_-(\mu)-\frac{3}{2}\xi_tC_-^{EW}(\mu)]\bar O_-
\biggr\}\nn
\mlab{14.34}
\ea
where ${\cal H}_I$ denotes the Hamiltonian which transforms as isospin I.
The operators above are defined as:
$$
\begin{array}{lll}
\bar O_\pm=\half(O^u_\pm-O^d_\pm),&~~~~~&
\hat O_\pm=\half(O^u_\pm+O^d_\pm),\\
O^c_\pm=\half(O_2^c\pm O_1^c),&~~~~~&
C_\pm^{EW}=C_{10}\pm C_9.
\end{array}
$$

In studying $B\to K\pi~\mbox{and~}\pi\pi$ decays, it is important to classify
final states in terms of strong interaction eigenstates, \ie~ isospin states.
This will allow us to take in to account of all rescattering effects which 
have been discussed extensively in the literature.
$$
\begin{array}{rll}
A(B^+\to\pi^+K^0)&=B_\half+A_\half+A_{\frac{3}{2}}&=P+A-\frac{1}{3}P_{EW}^c\\
-\sqrt{2}A(B^+\to\pi^0K^+)&=B_\half+A_\half-2A_{\frac{3}{2}}&
=P+T+C+A+\frac{2}{3}P_{EW}^c+P_{EW}\\
-A(B^0\to\pi^-K^+)&=B_\half-A_\half-A_{\frac{3}{2}}&=P+T+\frac{1}{3}P_{EW}^c\\
\sqrt{2}A(B^0\to\pi^0K^0)&=B_\half-A_\half+2A_{\frac{3}{2}}&=P-C
-\frac{2}{3}P_{EW}^c-P_{EW}
\end{array}
$$
where $A_I=\bra (K\pi)_I|{\cal H}_1|B\ket$ and 
$B_\half=\bra (K\pi)_\half|{\cal H}_0|B\ket$.
We have also given the decay amplitudes in terms of amplitudes classified by
Feynman graph structure: tree graph (T), color suppressed tree graph (C), annihilation graph (A), penguin graph (P), electroweak penguin graph ($P_{EW}$), and 
color suppressed electroweak penguin graph ($P_{EW}^c$).

These decays together with their charge conjugate version constitue 8 physical
observables. Unlike $K\to\pi\pi$ decays Watson's theorem cannot be applied, and
we cannot say much about final state interaction phases for these amplitudes.
We thus write \cite{nr}:
\be
A_i=a_ie^{i\alpha_i}+b_ie^{i(\beta_i+\phi_3)}~~~~~~~~i=1,2,3.
\mlab{11}
\ee
Here we separate the contributions which are proportional to 
$\xi_u\equiv |\xi_u|e^{i\phi_3}$ from 
those proportional to $\xi_c$ and $\xi_t$. $\alpha_i$ and $\beta_i$ are
final state strong interaction phases.
It is trivial to write $a_i$ and $b_i$ in terms of matrix elements of the effective Hamiltonian \mref{14.34}. Note that there are 12 independent parameters in
\mref{11}, and only 8 decay rates for $B\to K\pi$. 

In terms of matrix elements of the Hamiltonian, we have
\be
B_{\half}\equiv P;~~~
A_\frac{3}{2}=\bar C_+\bar Q_+^\frac{3}{2}+\bar C_-\bar Q_-^\frac{3}{2};~~~
A_\frac{1}{2}=\bar C_+\bar Q_+^\frac{1}{2}+\bar C_-\bar Q_-^\frac{1}{2},
\mlab{c16}
\ee
where 
\be 
\bar C_\pm=\xi_u C_\pm-\frac{3}{2}\xi_tC_\pm^{EW}=|\xi_u| C_\pm(e^{i\phi_3}-\delta_\pm),
\ee 
and $\bar Q^I_\pm$ is an
appropriate matrix element $\mat{(K\pi)_I}{\bar O_\pm}{B}$.
We also record
\be
P=\hat C_+\hat Q_+^\frac{1}{2}+\hat C_-\hat Q_-^\frac{1}{2}+\xi_c[C_+Q^c_++C_-Q_-^c]+\xi_t\sum_{i=3}^6C_iQ_i
\mlab{gro}
\ee
where 
$$
\hat C_\pm=|\xi_u| C_\pm(e^{i\phi_3}-\frac{1}{3}\delta_\pm),~~
\hat Q_\pm^\frac{1}{2}=\mat{(K\pi)_\half}{\hat O_\pm}{B},~~\mbox{and}~~
Q^c_\pm=\mat{(K\pi)_\half}{\hat O_\pm^c}{B}.
$$

So far, we have been quite general. Now, we shall go on to discuss the the hadronic matrix elements.
What do we know about these matrix elements? Over the years, we have learned quite a bit. In particular, we have learned that classifying operators in terms of their topology, A, C, T, P, $etc.$, gives us fairly accurate intuition in guessing the size of the matrix elements. For example, we guess that A, the annihilation graph, should be
quite small compared to T because it is suppressed by the small probability that the spectator quark and b quark come within the 
range so that they can annihilate. Similarly, C is suppressed compared to T because of the color factor. These statements imply definite relationships between hadronic matrix elements which appear in \mref{c16}. These
relations should be checked experimentally. But, for the time being, we shall
proceed and ask if these conventional wisdom would allow us to determine
$\phi_3$, the KM phase. 
The first simplification is 
$A(B^+\to\pi^+K^0)=B_\half+A_\half+A_{\frac{3}{2}}\approx |P|e^{i\phi_P}$ if A and $P_{EW}$
is negligible compared to P.
The second simplification is that $\bar Q_-$ transforms like a $\Delta I=0$ operator in the limit of U spin symmetry. So, $\bar Q_-^\frac{3}{2}$ vanishes in the SU(3) limit and is 
proportional to SU(3) breaking interaction. For our purpose, we neglect it. Then
$-3A_\frac{3}{2}=|P|\epsilon_{3/2}e^{i\phi_{3/2}}(e^{i\phi_3}-\delta_+)$,
where $|P|\epsilon_{3/2}=-3|\xi_u| C_+\bar Q_+^\frac{3}{2}$.
These considerations make analysis of $B^+\to K^+\pi^0$ simple:
\be
-\sqrt{2}A(B^+\to K^+\pi^0)=A(B^+\to K^0\pi^+)-3A_\frac{3}{2}
\ee
Neubert considers \cite{nr2}
\be
R_*^{-1}=\frac{2[{\rm Br}(B^+\to K^+\pi^0)+{\rm Br}(B^-\to K^-\pi^0)]}
{{\rm Br}(B^+\to K^0\pi^+)+{\rm Br}(B^-\to K^0\pi^-)}=1-2\epsilon_{3/2}\cos\Delta\phi(\cos\phi_3-\delta_+)
\mlab{Rstar}
\ee
where $\Delta\phi=\phi_\frac{3}{2}-\phi_P$. 
Fleischer considers \cite{flei2}
\be
R=\frac{[{\rm Br}(B^0\to K^+\pi^-)+{\rm Br}(\overline B^0\to K^-\pi^+)]}
{{\rm Br}(B^+\to K^0\pi^+)+{\rm Br}(B^-\to \overline K^0\pi^-)}.
\ee
The decay $B^0\to K^+\pi^-$ is bit more complicated because we have to
confront the contribution from $A_\half$. It involves two complex amplitudes.
He suplements the complexity by also considering
\be
A_0=\frac{{\rm Br}(B^0\to K^+\pi^-)-{\rm Br}(\overline B^0\to K^-\pi^+)]}
{{\rm Br}(B^+\to K^0\pi^+)+{\rm Br}(B^-\to \overline K^0\pi^-)}
\ee
Detailed numerical analysis indicates that both of these methods are useful
for determining $\phi_3$. 
In this discussion, I had to simplify the 
problem in order to present the essence. I refer the reader to the original 
contribution for complete analysis. 
It is clear that their contributions lead to much progress but much more
work is necessary along this direction. 
For example, only subset of $B\to K\pi$ has been 
considered. There are 8 of them altogether!

\section{SUSY in $B$ and $K$ decays}
Predictions of the minimal supersymmetric theories (MSSM)
is essentially same as those of the SM. If nature has
chosen MSSM, we will not learn anything new from experiments
on $B$ and $K$ decays. We should not be too discouraged by this
though, as it is likely that she has chosen a theory which
is more elegant than the MSSM. But, as long as we can not
specify which theory nature has chosen, it is not an easy task to analyze its
predictions. There are as many as 124 parameters in a non-
minimal SUSY, and perhaps even more.
Because $B$ and $K$ decays give stringent restrictions
on flavor changing neutral current strengths, we shall
focus general predictions of FCNC processes of a non-minimal
SUSY - mostly penguin effects.

In SUSY, there is a bosonic partner for each helicity of
a quark.  Here we begin by examining a $6\times 6$ squark mass matrix
of the MSSM.
\be
{\bf \tilde M}^2_D=\pmatrix{
{\bf \tilde M}_{DLL}^{\rm tree}{\bf \tilde M}_{DLL}^{\rm tree\dagger}
+c_1{\cal M}_U{\cal M}_U^\dagger
&Am_{3/2}{\cal M}_D(1+\frac{c_2}{M_W^2}
{\cal M}^\dagger_U{\cal M}_U)\cr
A^*m_{3/2}(1+\frac{c_2}{M_W^2}
{\cal M}^\dagger_U{\cal M}_U){\cal M}^\dagger_D
&{\bf \tilde M}_{DRR}^{\rm tree\dagger}{\bf \tilde M}_{DRR}^{\rm tree}}
\mlab{massmw}
\ee
where
\ba
{\bf \hat M}^2_{DLL}&=&\left( m^2_{3/2}+(v_1^2-v_2^2)
\left(\frac{{g'}^2}{12}- \frac{g^2}{4}\right)\right){\bf 1}+
({\cal M}^{\rm diag}_D)^2\nn
{\bf \hat M}^2_{DRR}&=&\left(m^2_{3/2}+ 
(v_1^2-v_2^2)\frac{{g'}^2}{6}\right){\bf 1} +({\cal M}^{\rm diag}_D)^2\nn
{\bf \hat M}^2_{DLR}&=&\left(|A|m_{3/2}+\mu^*\frac{v_1}{v_2}\right)
{\cal M}^{\rm diag}_D.
\mlab{massplanck}
\ea
${\bf 1}$ is a $2\times 2$ unit matrix, ${\cal M}_{D,(U)}$ is a mass matrix of a $D(U)$ quark, and 
${\cal M}^{\rm diag}_{D,(U)}$ is a corresponding diagonalized matrix.
Note that there new FCNC effects from additional flavor mixing among squarks.
Others are standard MSSM parameters. To go from MSSM to a more general theory, let us identify
\ba
\Delta_{LL}^2&=&c_1{\cal M}_U{\cal M}_U^\dagger\nn
\Delta_{LR}^2&=&Am_{3/2}{\cal M}_D
\left( 1+\frac{c_2}{M_W^2}
{\cal M}^\dagger_U{\cal M}_U \right) \nn
\Delta_{RL}^2&=&A^*m_{3/2}\left( 1+\frac{c_2}{M_W^2}
{\cal M}^\dagger_U{\cal M}_U\right) {\cal M}^\dagger_D. 
\mlab{DELTAAB} 
\ea 
and generalize $\delta_{LL}=\frac{\Delta_{LL}}{\tilde m}$, 
$\delta_{LR}=\frac{\Delta_{LR}}{\tilde m}$, 
$\delta_{RL}=\frac{\Delta_{RL}}{\tilde m}$
, and $\delta_{RR}=\frac{\Delta_{RR}}{\tilde m}$, where
$\tilde m$ is an average squark mass, 
as new arbitrary dimensionless parameters. 
We then study experimental constraints on $\delta$ ignoring all
theoretical prejudice.

This analysis has been discussed in detail by Masiero. 
Bounds on $\delta$ has been obtained from experimental
information on various FCNC processes. For 
an average squark mass and gluino mass of $500$GeV, 
the bounds on $\delta$ ranges from $10^{-1}$ to
$10^{-3}$. It sould be noted that the neutron edm
gives a bound of $\Im \left(\delta_{LR}\right)_{11}\sim 10^{-6}$
and it is natural to assume that other components of $\delta$ 
are of the same order of magnitude. If this is the case, it may be 
difficult of SUSY effects to show up in B and K decays.

\section{B decays to $\eta K(K^*)$, $\eta'K(K^*)$, and
$\eta'X_s$}
CLEO has observed \cite{cleoeta}
\ba
{\rm Br}(B\to \eta'K^+)&=&\left(6.5{{+1.5}\atop{-1.4}}\pm 0.9\right)\times 10^{-5},\nn
{\rm Br}(B\to \eta'K^0)&=&\left(4.7{{+2.7}\atop{-2.0}}\pm 0.9\right)\times 10^{-5}.\nn
\ea
These branching ratios are surprisingly large. Du has presented
a review of work in progress to undersand these large
branching ratios. It is likely that these branching ratio is large because
of gluonic content of $\eta'$.

Among various suggestions, a particularly
interesting one is that of Soni and Atwood. They attempt
to compute $\eta'\to$ glue glue by considering triangle
anomaly \cite{as}. They then estimate $b\to\eta'$ gluon.
We should note, however, that major contribution to this decay comes from
off shell gluon. Thus the validity of the anomaly calculation is questionable
at best. A universal characteristic of all the work presented by Du is that
each author picks up their favorite diagram and estimates its contribution.
Nature does not work this way. They have to consider all possible diagrams
and sum them up. Clearly global analysis is urgently needed.
Also, there are large amount of data on the gluonic content of $\eta'$.
Such global analysis must be consistent with the previously known properties
of $\eta'$.
\section{A new calculation for direct \cp~violation:
 $\frac{\epsilon'}{\epsilon}$}
Ever since the discovery of \cp~violation, experimentalists have
been looking for an evidence of direct \cp~violation.
Now that the result of NA31 has been confirmed by E832,
the direct \cp~ violation has been experimentally established.

The challenge for theorists is to extract physics from
the new value of $\frac{\epsilon'}{\epsilon}$. This
is not an easy task. Before we compute \cp~violating
amplitudes for $K\to\pi\pi$ decay, we have to demonstrate
that we understand \cp~conserving decay amplitudes.
This means that we need to understand the $\Delta I=\half$
rule. 
At the moment there is no clear understanding of this rule.
So, one choice is to obtain hadronic matrix elements
 from data \cite{buras}. If the
$\Delta I=\half$ rule is due to some new physics, this procedure
will miss the new physics. Clearly, this is not satisfactory.
We want to compute everything from basic principles. 
The approach taken by Paschos is an attempt along this direction.

Let us start from the defining equation:
\be
\epsilon ^{\prime} =e^{i(\delta_2-\delta_0)}
\frac{1}{2\sqrt{2}}\omega  
\left( \frac{\overline A_0}{A_0} - \frac{\overline A_2}{A_2}\right) .
\mlab{6.107}
\ee
where $A_I=\mat{(\pi\pi)_I}{{\cal H}}{K}$ and $\omega=|A_2/A_0|\sim .05$
and $\delta_I$ is the $\pi\pi$ phase shift for isospin $I$ channel.

In terms of operators in the effective Hamiltonian ${\cal H}$, 
\be
\frac{\epsilon'}{\epsilon}
=-ie^{i(\delta_2-\delta_0)}\times 10^{-4}
\left(\frac{\Im\tau}{10^{-4}}\right)
r\left(\sum_{i}y_i\bra Q_i\ket _0-\frac{1}{\omega}\sum_{i}
y_i\bra Q_i\ket _2\right)
\mlab{epe8}
\ee
where $r=\frac{G_F\omega}{|2\epsilon|\Re A_0}=336~{\rm GeV}^{-3}$; 
$y_i$ is the imaginary part of Wilson coefficients;
$\bra Q_i\ket _I$ is a hadronic matrix element
$\mat{(\pi\pi)_I }{0_i}{K}$; 
$\tau=-\frac{{\bf V}^*_{ts}{\bf V}_{td}}{{\bf V}^*_{us}{\bf V}_{ud}}$.

In this workshop, Paschos described computation of hadronic
matrix elements $\bra Q_i\ket _I$ based on 
the $\frac{1}{N_C}$ expansion. 
They have obtained
\be
5\times 10^{-4}\leq \frac{\epsilon'}{\epsilon}\leq 22\times 10^{-4}
\ee
for $m_s=(150\sim 175)$MeV. Their prediction increases if smaller
values of $m_s$ is taken.

This is an on going study and much more work remains:
(I)The Wilson coefficients can be computed reliably only
at some large energy scale $\Lambda_c$. But hadronic matrix elements
can be computed only at low energy scale. So, there has to be some compromise.
They ave chosen $\Lambda_c\sim 800\mbox{MeV}$. Wilson coefficient
functions have large scale dependence in this region. When the coefficient
functions and hadronic matrix elements are combined, the result for
$\frac{\epsilon'}{\epsilon}$ should not have the scale dependence. This
has to be studied carefully.
(II) It is necessary to
demonstrate that $K\to\pi\pi$ decay can be understood in this framework.
Indeed, their result for $K^+\to\pi^+\pi^0$ decay, the $\Delta I=\frac{3}{2}$
channel, is consistent with experiment. But it is necessary to understand
the $\Delta I=\frac{1}{2}$ amplitude. Personally, I am skeptical toward
a claim that the $\Delta I=\half$ rule can be understood within the frame
work of $\frac{1}{N_C}$ expansion. 
\section{Summary}
We have tried to have extended discussions on penguins. 
We tried to understand how we might extract $\phi_2$ and $\phi_3$ from data.
We don't think there is one best method to extract these angles.
It is an experiment driven field, and time will tell. We tried to understand 
$\frac{\epsilon'}{\epsilon}$ from basic principles of the SM. There is much more
work to be done along this direction. We tried to understand $B\to\eta'+X$ decays. This also requires more work. Seeing effects of SUSY in $B$ and $K$ decays
is an exciting possibility. We have to keep on searching.

\sk
\centerline{Acknowledgements}
This work has been supported in part by Grant-in-Aid for Special Project 
Research 
(Physics of \cp~violation). Comments from J. L. Rosner was helpful in finalizing the manusacript. I wish to express my gratitude to the organizers, 
C. A. Dominguez and R. D. Viollier for their effort in organizing the 
workshop, and for their hospitality.


\section*{References}
\begin{thebibliography}{99}
\bibitem{cleo}J. Alexander, talk presented at the 29th ICHEP, Vancouver, B. C. 
Canada, (1998).
\bibitem{E832}
http://fnpx03.fnal.gov/experiments/ktev/ktev.html
\bibitem{E731}
E731: L. K. Gibbons \etal, \PRL{70}{1203}{1993} .
\bibitem{NA31}
NA31 : G. D. Barr \etal, \PL{B317}{233}{1993} .
\bibitem{gro2}See M. Gronau, this proceedings.
\bibitem{srs}J. L. Rosner, A. I. Sanda, and M. Schmidt
{\it Proc. Fermilab Workshop on High Sensitivity Beauty Physics
at Fermilab}, A. J. Slaughter, N. Lockyer and M. Schmidt (eds.),  1987.
See also,
C. Hamzaoui, J. L. Rosner and A. I. Sanda, same proceeding.
\bibitem{gro2}M. Gronau and D. Pirjol hep-ph/9902482.
\bibitem{charl}J. Charles LPTHE-ORSAy 98-35 to appear in the Physical Review.
\bibitem{nr}M. Neubert and J. L. Rosner, \PL{B441}{403}{1998} .
\bibitem{flei}R. Fleischer, {\it Eur. Phys. J.}{\bf C}(1998) DOI 10.1007/s100529800919[hep-ph/9802433].

\bibitem{nr2}M. Neubert, this workshop; M. Neubert and J. L. Rosner, \PRL{81}{5076}{1998} .
\bibitem{flei2}R. Fleischer, this workshop.

\bibitem{cleoeta}CLEO Collaboration  R. H. Behrens \etal, \PRL{80}{3710}{1998} .
D. Atwood and A. Soni hep-ph/98043
\bibitem{as}D. Atwood and A. Soni hep-ph/98043
\bibitem{buras}
G. Buchalla, A. J. Buras and M. E. Lautenbacher, \NP{B370}{69}{1992} ;
A. J. Buras, M. Jamin and M. E. Lautenbacher, \NP{B408}{209}{1993} .

\end{thebibliography}
\end{document}